\documentclass{article}
\usepackage{epsfig}
\def\be{\begin{eqnarray}}
\def\ee{\end{eqnarray}}

\hoffset= - 1.0in

\begin{document}

\hfill ITEP/TH-35/07 \\ \\

\centerline{\Large{ Universal Mandelbrot Set }}

\centerline{\Large{ as a Model of Phase Transition Theory }}

\bigskip

\centerline{Andrey Morozov\footnote{Andrey.Morozov@mail.itep.ru}}

\bigskip

\centerline{\it Moscow State University}

\centerline{\it Institute of Theoretical and Experimental Physics}

\bigskip

\bigskip

\centerline{ABSTRACT}

\bigskip

{\footnotesize The study of Mandelbrot Sets (MS) is a promising new
approach to the phase transition theory. We suggest two improvements
which drastically simplify the construction of MS. They could be
used to modify the existing computer programs so that they start
building MS properly not only for the simplest families. This allows
us to add one more parameter to the base function of MS and
demonstrate that this is not enough to make the phase diagram
connected.}

\bigskip

The problem of stability of time evolution is one of the most
important in physics. Usually one can make the motion stable or
unstable by changing some parameters which characterize Hamiltonian
of the system. Stability regions can be represented on the phase
diagram and transitions between them are described by catastrophe
theory \cite{Catstrophe}, \cite{Arnold}. It can seem that a physical
system or a mechanism can be taken from one domain of stability to
any other by continuous and quasi-static variation of these
parameters, i.e. that the phase diagram is connected. However
sometimes this expectation is wrong, because domains of stability
can be separated by points where our system is getting totally
destroyed. Unfortunately today it is too difficult to explore the
full phase diagram for generic physical system with many parameters.
Therefore, following \cite{LL6}, it was proposed in \cite{DM} to
consider as a simpler model the discrete dynamics of one complex
variable \cite{Milnor}-\cite{Shabat}. The phase diagram in this case
is known as Universal Mandelbrot Set (UMS). MS is a well-known
object in mathematics \cite{Wiki}-\cite{Fe}, but its theory is too
formal and not well adjusted to the use in the phase transition
theory. The goal of this paper is to make MS more practical for
physical applications.

\section{Structure of MS\label{MSstruct}}

\begin{figure}\center
\epsfxsize 350pt \rotatebox{0}{\epsffile{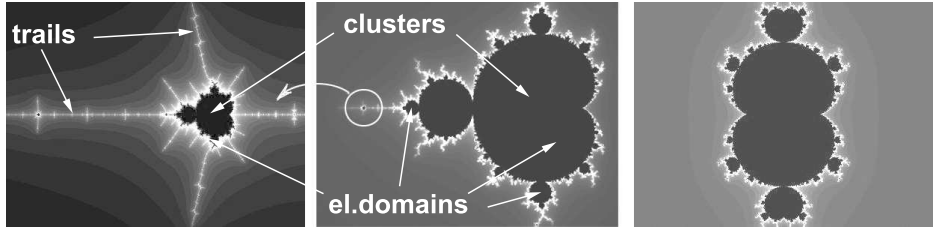}}
\caption{\footnotesize{The simplest examples of Mandelbrot sets
$MS(x^2+c)$ and $MS(x^3+c)$, constructed by Fractal Explorer
\cite{Fe}. The picture explains the terms``clusters'',``elementary
domain'' and ``trails''. It is difficult to see any clusters except
for the central one in the main figure, therefore one of the smaller
clusters is shown in a separate enlarged picture to the left.}}
\label{F5}
\end{figure}

First of all we remind the definition of MS and UMS from \cite{DM},
which different from conventional definition in mathematical
literature, see s. \ref{fastm} below. Mandelbrot Set (MS) is a set
of points in the complex $c$ plane. MS includes a point $c$ if the
map $x\rightarrow f(x,c)$ has stable periodic orbits. As shown in
Fig.\ref{F5} MS consists of many clusters connected by trails, which
in turn consist of smaller clusters and so on. Each cluster is
linear connected and can be divided into elementary domains where
only one periodic orbit is stable. Different elementary domains can
merge and even overlap. Boundary of elementary domain of $n$-th
order, i.e. of a domain where an $n$-th order orbit is stable, is a
real curve $c(\alpha)$ given by the system:

\be \left\{\
\begin{array}{l}G_n(x,c)=0\\F'_n(x,c)+1=e^{i\alpha}\end{array}\right.
\label{norb} \ee with
$$F_n(x,c)=f^{\circ n}(x,c) - x$$
$$G_n(x,c)=\frac{F_n(x,c)}{\prod_m{G_m(x,c)}},\ n\vdots m$$
Indeed, when $G_n(x,c)$ vanishes then $x$ belongs to the orbit of
exactly the $n$-th order. This orbit is stable if $|\frac
{\partial}{\partial x}f^{\circ n}(x,c)|<1$, what implies the second
equation of (\ref{norb}). The solution of this system may give us
more than a single $n$-th order domain. Domains of different orders
merge at the points $c$ where

\be Resultant_x\Big(G_n(x,c),G_k(x,c)\Big)=0 \label{res} \ee Of
course two orbits and thereafter domains merge only if $n$ is
divisor of $k$, so it is reasonable to consider only
$Resultant_x(G_n,G_{mn})=0$, with $m=2,3\ldots$ and
$Discriminant_x(G_n)$ for $k=n$.

Physically MS is a phase diagram of discrete dynamic of one complex
variable. It is clear from Fig.\ref{F5} that one should distinguish
between three types of connectivity in different places of phase
diagram. The first is linear connectivity: the possibility to
connect any two points with a continuous line. The second type is
weak connectivity: it means that only a closure of our set has
linear connectivity. The third type we call strong connectivity: it
means that any two interior points are connected with a thick tube.

Entire MS on Fig.\ref{F5} is weakly\footnote{MS is usually claimed
to be locally connected \cite{Wiki}, i.e. any arbitrary small
vicinity of a point of MS contains a piece of some cluster. In our
opinion weak connectivity is another feature, especially important
for physical applications.} but not linearly connected and its
clusters are linearly, but not strongly connected. Universal
Mandelbrot Set (UMS) is unification of MS of different
$1_c$-parametric families. When we rise from MS to UMS we add more
parameters to the base function. Thus entire UMS could become
strongly connected, but it is unclear whether this really happens.

\section{Simplification of the resultant condition (\ref{res})\label{simplres}}

In this section we prove that the resultant condition (\ref{res})
can be substituted a by much simpler one: \be
Resultant_x\Big(G_n(x),(F'_n+1-e^{i\alpha})\Big)=0,\ \ \ \ \
\alpha=\frac {2\pi}{m}k  \label{res2}\ee To prove (\ref{res2}) it is
enough to find the points where $Resultant_x(F_n,
\frac{F_{nm}(x)}{F_n(x)})=0$. Then:
$$Resultant_x(F_n,\frac{F_{nm}(x)}{F_n(x)})=
Resultant_x(F_n,\frac{F'_{nm}(x)}{F'_n(x)})=0$$ By definition of
$F(x)$:
$$F_{nm}(x)=F_{n(m-1)}\Big(F_n(x)+x\Big)+F_n(x)$$
Then:
$$\frac{F'_{nm}(x)}{F'_n(x)}=
\frac{F'_{n(m-1)}(x)}{F'_n(x)}(F'_n(x)+1)+1\
\stackrel{(\ref{norb})}{=}\
\frac{F'_{n(m-1)}(x)}{F'_n(x)}e^{i\alpha}+1= \ldots$$
$$=(\ldots((e^{i\alpha}+1)e^{i\alpha}+1)e^{i\alpha}+1)\ldots)+1=
\sum^{m-1}_{l=0}{e^{li\alpha}}=
\frac{e^{mi\alpha}-1}{e^{i\alpha}-1}$$ Thus (\ref{res}) implies that
$e^{mi\alpha}-1=0$ and therefore $\alpha = \frac{2\pi k}{m}$

This theorem is a generalization of a well known fact for the
central cardioid domain of $MS(x^2+c)$ (see for example
\cite{Wiki}). This also provides a convenient parametrization of
generic MS.


\section{A fast method of MS simulation\label{fastm}}

Historically MS was introduced in a different way from section
\ref{MSstruct}. We call it $\widetilde{MS}$. It depends not only on
the family of functions, but also on a point $x_0$. If $c$ belongs
to the $\widetilde{MS}(f,x_0)$ then
$$\lim_{n\rightarrow\infty}f^{\circ n}(x_0)\neq\infty$$
In the literature one usually puts $x_0=0$ independently of the
shape of $f(x)$. Such $\widetilde{MS(f,0)\neq}MS(f)$ except for the
families like $f=x^a+c$. Existing computer programs \cite{Fe}
generate $\widetilde{MS}(f,0)$, and can not be used to draw the
proper $MS(f)$. Fortunately there is a simple relation: \be
MS(f)=\bigcup_{x_{cr}}\widetilde{MS}(f,x_{cr})\label{hyperconj}\ee
where union is over all critical points of $f(x)$, $f'(x_{cr})=0$.
Equation (\ref{hyperconj}) is closely related to hyperbolic and
local connectivity conjectures \cite{Wiki}. It is also equivalent to
the following two statements about the phase portrait in the complex
x plane:

(I) If  $\lim_{l\rightarrow\infty}f^{\circ l}(x_{cr})\neq\infty$
then there is a stable periodic orbit {\cal O} of finite order which
attracts $x_{cr}$. It implies that
$$MS(f)\supseteq\bigcup_{x_{cr}}\widetilde{MS}(f,x_{cr})$$

(II) If {\cal O} is a stable periodic orbit, then a critical point
$x_{cr}$ exists, which is attracted to {\cal O}. This implies that
$$MS(f)\subseteq\bigcup_{x_{cr}}\widetilde{MS}(f,x_{cr})$$

The statement (I) says that if $\lim_{l\rightarrow\infty}f^{\circ
l}(x_{cr})\neq\infty$ then this limit exists and is a stable orbit
with finite period, i.e. that there are no such things as strange
attractors in discrete dynamics of one complex variable. This
statement is unproved but we have no counter-examples.

The statement (II) is much easier. If $x_0$ is a stable fixed point,
then it is surrounded by a disk-like domain, where $|f'(x)|<1$. Its
boundary is parametrized by $f'(x)=e^{i\alpha}$ and inside this area
there is a point where $f'(x)=0$, i.e. some critical point $x_{cr}$
of $f$. It is important that this entire surrounding of $x_0$ -- and
thus this $x_{cr}$ -- lie inside the attraction domain of $x_0$:
$$|f(x_{cr})-f(x)|<|x_{cr}-x|$$
i.e. we found $x_{cr}$ which is attracted to $x_0$. This argument
can be easily extended to higher order orbits and can be used to
prove (II).

\begin{figure}\center
\epsfxsize 350pt \rotatebox{0}{\epsffile{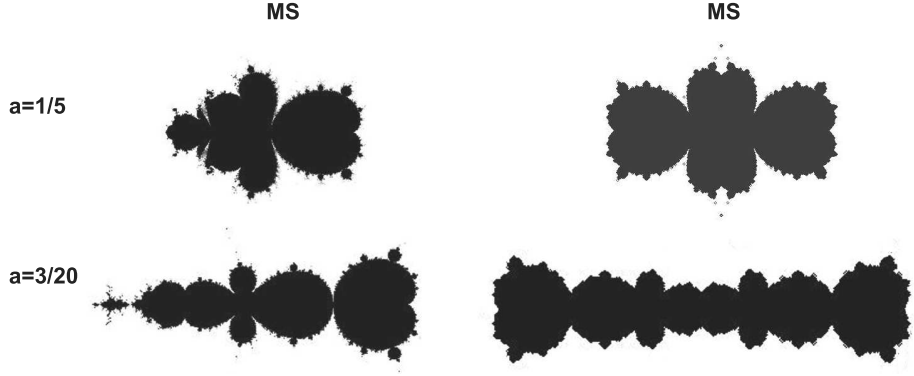}}
\caption{\footnotesize{The families of Mandelbrot Sets
$MS\left(a\cdot x^3+(1-a)\cdot x^2+c\right)$ at $a=1/5$ (the upper
one) and $a=3/20$. The left pictures show the wrong Mandelbrot Set
$\widetilde{MS}(x=0)$, generated by Fractal Explorer \cite{Fe}
.According to (\ref{hyperconj}), the proper Mandelbrot Set is
$MS=\widetilde{MS}(0)\bigcup \widetilde{MS}(-\frac{2(1-a)}{3a})$,
and it is shown in the right pictures. In this particular case,
because of the $Z_2$ symmetry \cite{DM}, the two different
$\widetilde{MS}(x=0)$ are related by reflection w.r.t. axis
$\textrm{Re}(c)=\frac{(a-1)(a+2)(2a+1)}{27a^2}$.}} \label{FEim}
\end{figure}

Equation (\ref{hyperconj}) leads to a simple upgrade of programs,
which construct MS. Fig. \ref{FEim} demonstrates a result of this
improvement.

\section{A way from $MS(x^3+c)$ to $MS(x^2+c)$}

\begin{figure}\center
\epsfxsize 450pt \rotatebox{0}{\epsffile{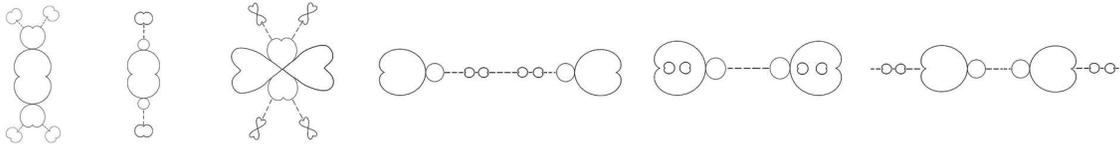}}
\caption{\footnotesize{Schematic picture, showing the central
cluster and the first non-trivial (order-2) clusters of $MS(a\cdot
x^3+(1-a)\cdot x^2+c)$ at different values of $a$. The small
clusters are artificially enlarged to make them visible. Their
shapes imitate that of the central cluster, in accordance with
\cite{DM2}.All clusters move and pass through each other (i.e. the
corresponding orbits are simultaneously stable at the same values of
$c$, but remain different, the corresponding resultants do not
vanish). This means that this $2_C$-parametric section of UMS is
still not linear connected.}} \label{fivefig}
\end{figure}

As application of our results in ss.\ref{simplres} and \ref{fastm}
we consider the $2_C$-parametric section of UMS for the family
$f(x)=a\cdot x^3+(1-a)\cdot x^2+c$, which interpolates between
$MS(x^3+c)$ at $a=1$ and $MS(x^2+c)$ at $a=0$. We extend
consideration of \cite{DM2}  to non-trivial second order clusters
which were beyond the reach of the methods used in that paper. The
result is schematically shown on Fig. \ref{fivefig}. We used the
Fast Method from section \ref{fastm} to draw the entire MS. And we
used simplification from section \ref{simplres} to find the merging
points of the first and the second domains and of the second and the
forth domains. The important outcome of this experiment is that
clusters, which were disconnected when $a=1$, remain disconnected
for all $a$ and go to the infinity when $a$ goes to zero. There is
no point between $a=0$ and $a=1$, where secondary clusters would
touch the central cluster. Thus adding one more parameter to the MS
does not make it linearly and strongly connected.

\section{Conclusion}

The Universal Mandelbrot Set is a representative model of
sophisticated phase structure in complicated physical systems.
However even this simpler model is still very difficult to explore
and understand. In this paper we proposed considerable
simplifications of the theory allowing to make the computer
experiments with the properly defined MS in a simple and efficient
fashion. This opens a way to attack the main puzzles such as the
nature of trails and connectivity of phase diagram.

\section*{Acknowledgements}

I appreciate discussions with V. Dolotin and A. Morozov and I thank
T. Mironova for help with the pictures. I acknowledge hospitality of
the ESI in Vienna, where part of this work was done. This work is
partly supported by Russian Federal Agency of Atomic Energy, by the
grant RFBR 07-01-00526 and the grant of support to the Scientific
Schools NSh-8065.2006.2.


\end{document}